\documentclass{nature}

\usepackage{amssymb}
\usepackage{epsfig}
\title{Bosonic excitations and electron pairing in an electron-doped cuprate superconductor}

\author{M. C. Wang$^{1}$, J. Xiong$^{2}$, H. S. Yu$^3$, Y. -F. Yang$^{3,4}$, S. N. Luo$^{1}$, K. Jin$^{3,4,\ast}$ \& J. Qi$^{1,2,\dagger}$}

\begin{document}

\maketitle

\begin{affiliations}
 \item The Peac Institute of Multiscale Sciences, Chengdu, Sichuan 610031, China
 \item School of Microelectronic and Solid-State Electronics, University of Electronic Science and Technology of China, Chengdu 610054, China
 \item Beijing National Laboratory for Condensed Matter Physics, Institute of Physics, Chinese Academy of Sciences, Beijing 100190, China
 \item Collaborative Innovation Center of Quantum Matter, Beijing 100190, China\\
 \normalsize{$^\ast$Corresponding author: kuijin@iphy.ac.cn}\\
 \normalsize{$^\dagger$Corresponding author: jbqi@uestc.edu.cn}

\end{affiliations}

\begin{abstract}
Superconductivity originates from the coupling between charge carriers and bosonic excitations of either phononic or electronic origin. Identifying the most relevant pairing glue is a key step towards a clear understanding of the unconventional superconductivity. Here, by applying the ultrafast optical spectroscopy on the electron-doped cuprates La$_{1.9}$Ce$_{0.1}$CuO$_{4\pm\delta}$, we discern a bosonic mode of electronic origin that has the strongest coupling with the charge carriers near $T_c$. We argue that this mode is associated with the two-dimensional antiferromagnetic spin fluctuations, and can fully account for the superconducting pairing. Our work may help to establish a quantitative relation between bosonic excitations and superconducting pairing in electron-doped cuprates, and pave the way for systematic exploration of superconductivity and other collective phenomena in all correlated materials.
\end{abstract}

In unconventional superconductors, such as high-$T_c$ cuprates, it has been a longstanding challenge to reveal the effective interaction of charge carriers (fermionic quasiparticles) with phonons or other bosonic excitations of electronic origin\cite{isotope_effect_doubt_phonon_pairing,arpes_spin_flucatuation_pairing,spin_fluctuations_Arpes}. In particular, huge efforts have been made in order to unravel the electronic excitations likely giving rise to the superconducting (SC) paring, such as charge, spin and orbital fluctuations\cite{nature_NCCO_spin_correlations,Fujita_2008_PRL,CDW_in_NCCO,CDW_in_YBCO,spin_fluctuations_Arpes,PLCCO_gap_STM_Nature,arpes_spin_flucatuation_pairing}. However, an unambiguous solution to the pairing issue remains obscure because it is extremely difficult to quantitatively identify the most relevant bosonic excitations and their related interactions near $T_c$, due to various types of excitations entangled in the energy domain\cite{IR_NCCO,IR_PCCO,IR_Bi2212,Arpes_NCCO_PG_induced_by_AFM}.   

Ultrafast optical spectroscopy provides a unique opportunity to directly probe the related interactions in the time domain\cite{Ultrafast_review}. Under non-equilibrium conditions, these bosonic excitations display distinct relaxation dynamics because of their different couplings to the charge carriers and may thereby be potentially disentangled. This technique has been extensively utilized in studying the SC gap, pseudogap and competing orders on high-$T_c$ superconductors\cite{pump_probe_YCBCO,pump_probe_LSCO,pump_probe_Bi2212,NCCO_pump_probe,Gedik_2016_arXiv,pump_probe_Tl2223}, but its application on unveiling the entangled bosonic excitations and their temperature evolution remains elusive\cite{Ultrafast_review,pump_probe_disentangle1_failed2,pump_probe_disentangle_failed1}.

In this work, by measuring the transient optical reflectivity change $\Delta R(t)/R$ as a function of temperature, we unravel different bosonic modes and investigate their peculiar relaxation dynamics in the electron-doped cuprates La$_{2-x}$Ce$_x$CuO$_{4\pm\delta}$ (LCCO). Unlike its hole-doped counterparts, where superconductivity is often intertwined with anomalous pseudogap, the electron-doped cuprates are relatively simple and believed to exhibit prominent antiferromagnetic (AFM) spin fluctuations (or correlations)\cite{e-doped_main,JinKui_nature}. In fact, our LCCO samples are optimally doped with Ce concentration of $x$=0.1, which sits near the AFM border in the temperature-doping ($T$-$x$) phase diagram. The temperature-dependent results enable us to identify with confidence a bosonic mode of electronic origin that is closely related to the two-dimensional (2D) AFM spin fluctuations. We discover an enormous enhancement of its interaction with fermionic quasiparticles as temperature decreases from high temperatures down to $T_c$. We demonstrate that this bosonic mode is of critical importance for the electron pairing and can fully account for the SC $T_c$.

\begin{figure}
	\centering
	\epsfig{figure=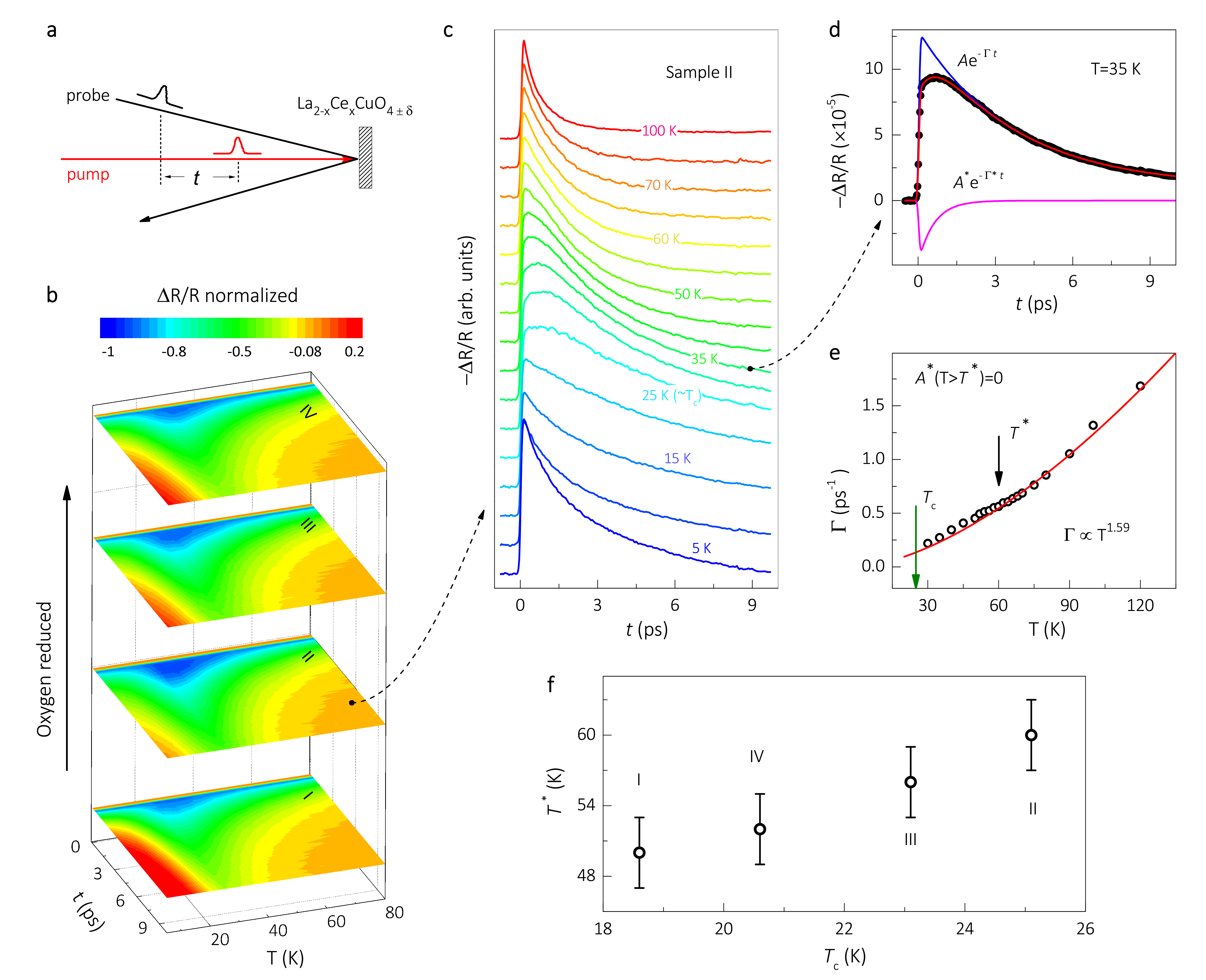,width=16cm}
	\caption{\label{fig1} \textbf{Temperature-dependent $\Delta R(t)/R$ in optimally doped La$_{1.9}$Ce$_{0.1}$CuO$_{4\pm\delta}$.} ({\bf a}) Schematic of optical pump-probe experimental setup. ({\bf b}) $\Delta R(t)/R$ as a function of temperature in four different samples I, II, III, and IV with different oxygen concentration $\delta$. ({\bf c}) Detail plot of $\Delta R(t)/R$ measured on sample II. ({\bf d}) A typical fitting of $\Delta R(t)/R$ at 35 K using two exponential decays shown by the red line. Two decay components are indicated by the blue ($Ae^{-\Gamma t}$) and green ($A^*e^{-\Gamma^* t}$) lines, respectively. ({\bf e}) The $\Gamma$ values at $T>T_c$ follows well a  fluence-independent power law :$\Gamma\propto T^{1.59}$. The black and green arrows indicate the positions of $T^*$ and $T_c$, respectively. ({\bf f}) $T_c$ increases monotonically with $T^*$ for the investigated samples.}
	
	\epsfig{figure=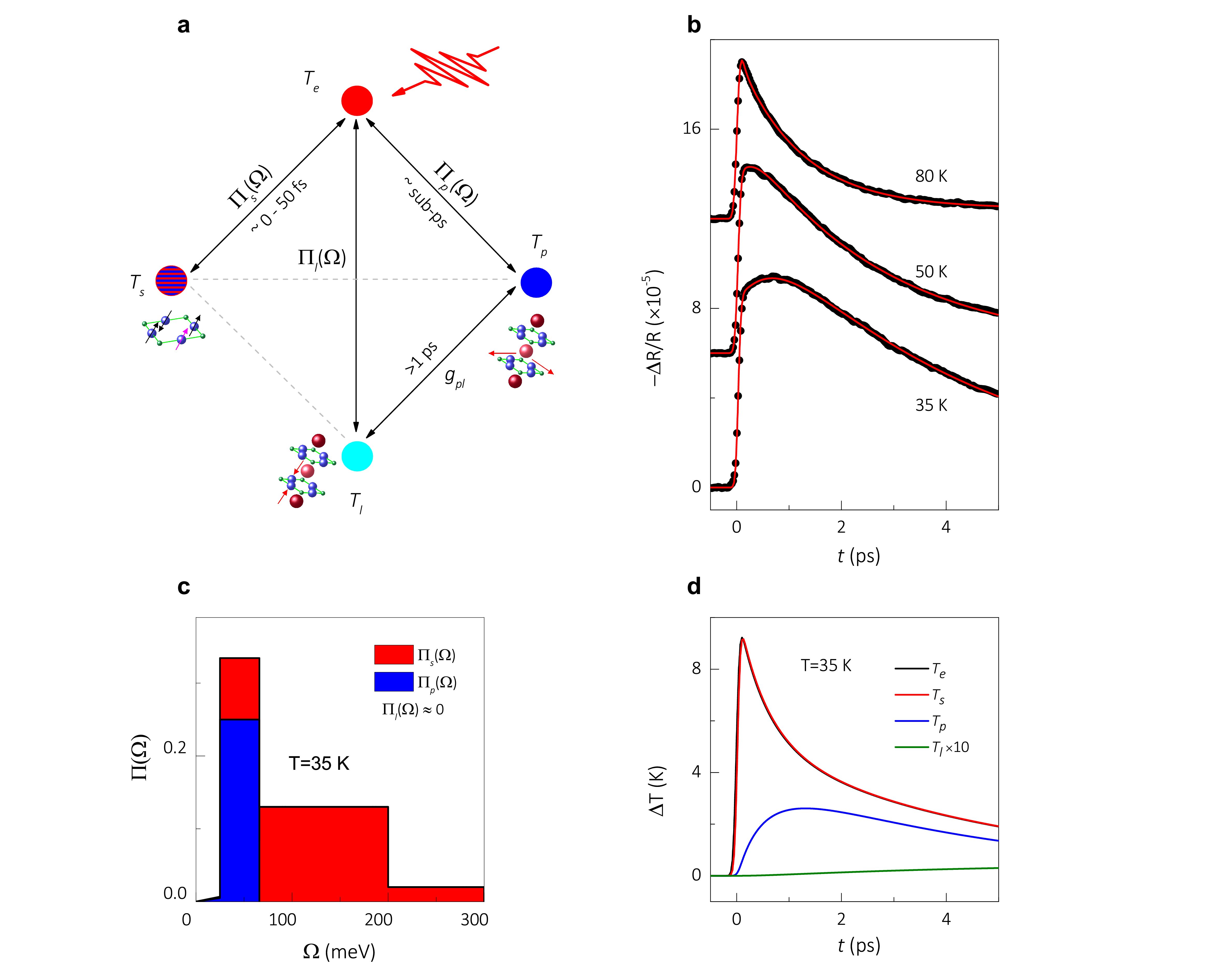,width=16cm}
	\caption{\label{fig2} \textbf{Typical fitting via the effective temperature model.} ({\bf a}) Energy relaxation between electrons ($T_e$), bosons with electronic origin ($T_s$), hot phonons ($T_p$), and rest of the lattice ($T_l$) are described by the effective temperature model (see Methods). Time evolution of $T_e$ and $T_j$ ($j=s,p,l$) are determined by the bosonic spectra function $\Pi_j(\Omega)$, and the anharmonic phonon decay, $g_{pl}$. The thermalization timescales are given by the fitted $T_e(t)$ and $T_j(t)$. ({\bf b}) Fitting results at several typical temperatures. ({\bf c}) The bosonic function $\Pi(\Omega)$ obtained by fitting $\Delta R/R$ at 35 K. The red and blue areas represent $\Pi_s(\Omega)$ and $\Pi_p(\Omega)$ associated with the electronic and phononic excitations, respectively. ({\bf d}) The corresponding $T_e(t)$ and $T_j(t)$ at 35 K, characterizing the electrons, bosons with electronic origin, hot phonons, and rest of the lattice, respectively.}

	\epsfig{figure=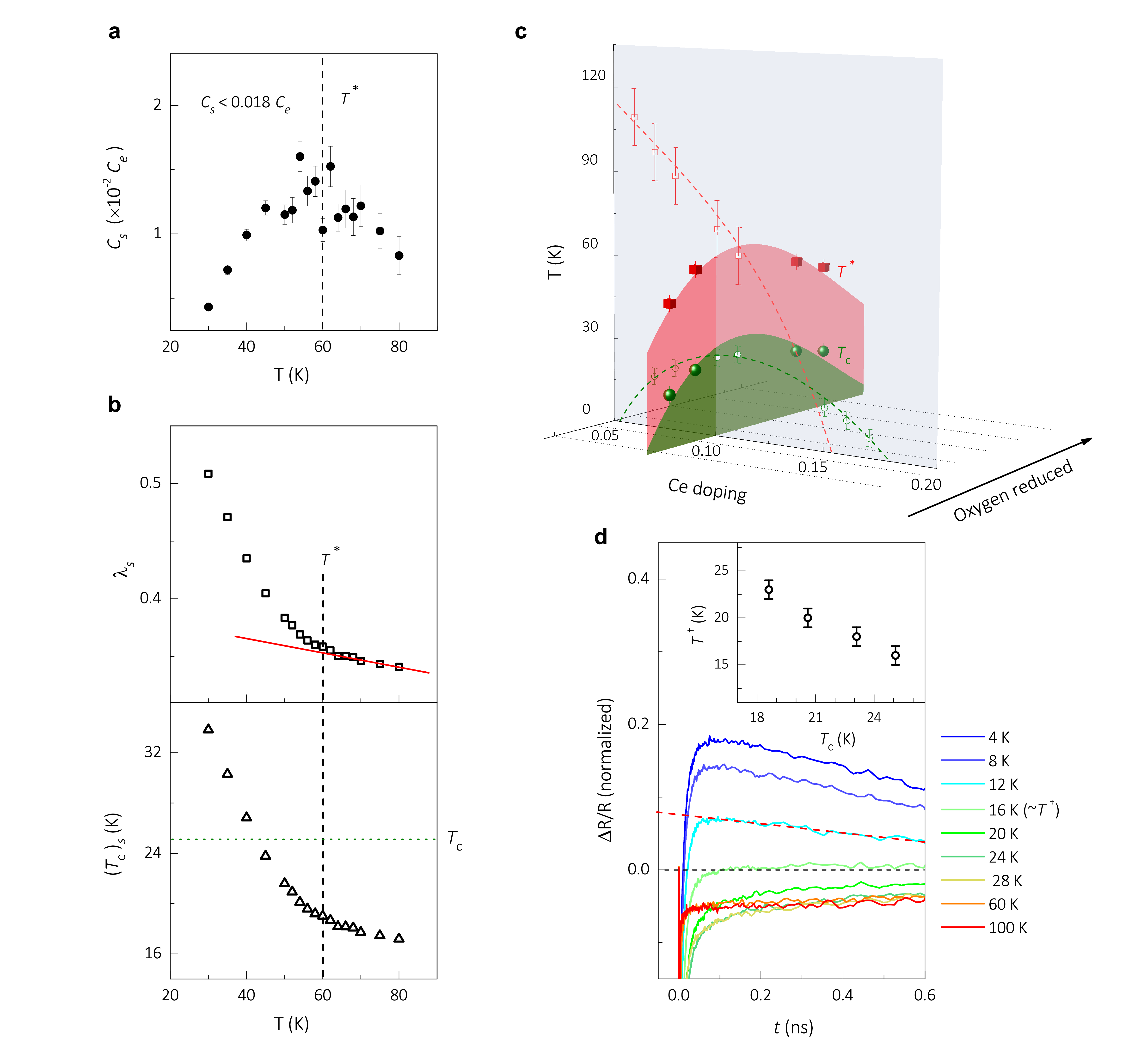,width=16cm}
	\caption{\label{fig3} \textbf{Temperature-dependent fitting results and the long relaxation processes in $\Delta R(t)/R$.}  ({\bf a}) The specific heat, $C_s$, as a function of temperature. $C_s$ reaches a maximum around $T^*$, indicated by the dash line. ({\bf b}) The electron-boson coupling constant $\lambda_s$ as a function of temperature (black open squares). The red solid line indicates the experimental data below the critical temperature $T^*$ strongly increase with decreasing temperature. The estimated maximum SC transition temperatures $(T_c)_s$ is shown by the black open triangles. Green dotted line indicates $T_c$(=25.1 K). Black dash-line shows the position of $T^*$. ({\bf c}) $T-x$ and $T-\delta$ phase diagram for LCCO. In $T-x$ phase diagram, the blue dashed line with open circles represent the SC boundary, while the red dashed line with open squares represent the AFM boundary estimated by the in-plane angular magnetoresistance measurements \cite{JinKui_AMR,JinKui_nature}. In $T-\delta$ phase diagram, the green area is the SC dome, while the red area, evaluated from this work, represents the 2D AFM spin fluctuation regime. ({\bf d}) $\Delta R(t)/R$ as a function of temperature on a long timescale for sample II. Red dashed line represents the long relaxation component extending into nanosecond regime. $T^\dagger$ is the critical temperature where amplitude of the long relaxation component flips sign. $T^\dagger$ as a function $T_c$ is shown in the inset.}
	
\end{figure}

Figure \ref{fig1}a shows the typical ultrafast optical pump-probe spectroscopy setup, where a femtosecond laser pulse excites the quasiparticles of the sample into a non-equilibrium state whose time-evolution is probed by monitoring the reflectivity change $\Delta R(t)/R$ of another time-delayed optical pulse. Figure \ref{fig1}b displays the obtained $\Delta R(t)/R$ as a function of temperature within $\sim$10 ps in four optimal-doped La$_{1.9}$Ce$_{0.1}$CuO$_{4\pm\delta}$ with different $\delta$. The signals in all samples are well consistent but behave distinctively in separate temperature regimes. Fig.~\ref{fig1}c plots $\Delta R(t)/R$ measured on sample II ($T_c$=25.1 K). The relaxation seems to take simple exponential decay at both high and low temperatures. However, a peculiar and relatively slow rising process is observed at intermediate temperatures between $\sim$25 K and $\sim$60 K as manifested by the bump-like behaviour following the initial instantaneous uprising. This type of signal corresponds to the dark blue region in Fig.~\ref{fig1}b, and seems to be quite common in cuprates\cite{Gedik_2016_arXiv,NCCO_pump_probe,pump_probe_LSCO}. Due to similarities between the four samples, all detailed analyses below are for data taken on sample II except where noted.

We mainly focus on the experimental data for $T>T_c$, measured with low pump fluence, so that we can ignore the complexity of interpreting the quasiparticle dynamics in the SC state\cite{bottleneck_main}. As seen in Fig.~\ref{fig1}d, $\Delta R(t)/R$, either with or without the bump-like feature, can be well fitted using a function of the form: $\Delta R(t)/R=Ae^{-\Gamma t}+A^*e^{-\Gamma^* t}$, where $A$ (or $A^*$) and $\Gamma$ (or $\Gamma^*$) are the amplitude and decay rate, respectively. Specifically, $A^*=0$ for $T>T^*$ ($T^*\simeq$60 K). Emergence of the non-zero $A^*$ component indicates the onset of new scattering channel between quasiparticles and bosons around $T^*$. Since this change occurs instantaneously after photoexcitation, we expect that it is caused by bosonic excitations that are strongly coupled with the non-equilibrium quasiparticles in the initial relaxation dynamics. Fig.~\ref{fig1}e plots the decay rate $\Gamma$ for $T>T_c$, where we find the data follow a well defined power law: $\Gamma\propto T^\alpha$ ($\alpha\simeq$1.59). This power law is derived by fitting the fluence-independent $\Gamma$ above $T^*$. Similar scaling has also been observed in the $T$-dependent resistivity\cite{JinKui_PNAS,JinKui_nature}, suggesting an intimate connection between the resistivity scaling and the coupling to the bosonic excitations. Here, low pump fluence is crucial as it helps yield fluence-independent parameters and enables us to extract accurate physical information using the effective temperature model described below. Because, at high pump fluences, $\Gamma$ below $T^*$ will obviously deviate from this power law\cite{Gedik_2016_arXiv}. In addition, we surprisingly notice that $T_c$ almost increases linearly with increasing $T^*$ for the investigated samples (see Fig.~\ref{fig1}f). Such result suggests that the bosonic excitations contributing to the new scatterings emerging below $T^*$ is closely related to the SC property.

In order to reveal the intrinsic mechanisms behind our observations, we adopt the effective temperature model to understand our experimental data\cite{Ultrafast_review,ETTM_Bi2212_Science} (see Methods). Based on the work by Giannett et al.\cite{ETTM_Bi2212_Science}, under the assumption of an instantaneous buildup of the Fermi-Dirac distribution after photoexcitation\cite{4TM_conference_quasi_thermal,TTM_non-thermal}, one can understand the temporal evolution of the non-equilibrium state as the energy exchange among the electron and boson reservoirs via the electron-boson and boson-boson scattering processes in the frame of effective temperature model. Fig.~\ref{fig2}a illustrates the energy exchanges and relaxation between four reservoirs in photoexcited systems: one for the charge carriers, and three for the bosonic excitations. Each reservoir is characterized by an effective temperature $T_e$ or $T_j$ ($j=s,p,l$). The subscript $e$ denotes the charge carriers. The bosonic excitations are classified by their electronic origin such as spin fluctuations ($s$), or phononic origin including hot phonons ($p$) and the rest of lattice excitations ($l$). Time-evolution of the effective temperatures is quantitatively connected with the interactions between charge carriers and bosonic modes. The interaction associated with each mode can be fully accounted by its linear contribution, $\Pi_j(\Omega)$, to the total bosonic spectral function $\Pi(\Omega)$\cite{ETTM_Bi2212_Science,Ultrafast_review}. Conventionally, $\Pi_s(\Omega)$ and $\Pi_{p,l}(\Omega)$ are also expressed as $I^2\chi(\Omega)$ and [$\alpha^2F(\Omega)$]$_{p,l}$. In principle, $T_e(t)$, $T_j(t)$ and $\Pi_j(\Omega)$, deciding the electronic self-energy $\Sigma(t, \Omega)$, can be derived from the time-dependent optical conductivity $\sigma(t, \omega)$ or reflectivity $R(t, \omega)$ using the extended Drude model\cite{ETTM_Bi2212_Science,4TM_conference_quasi_thermal}. Such relationship provides us the possibility to disentangle the electronic and phononic excitations, and quantitatively elucidate the temperature evolution of their interactions with the fermionic quasiparticles that leads to the electron pairing at $T_c$.

However, different from previous work done at room temperature\cite{ETTM_Bi2212_Science}, we find that during quasiparticle thermalization it is necessary to include the additional coupling term $g_{pl}$ describing the anharmonic decay to explain our $T$-dependent experiments in low temperature regime (see Methods), as illustrated in Fig.~\ref{fig2}a. Therefore, hot phonons play an indispensable and dominant scattering medium for the excited quasiparticles to equilibrate with the lattice. In fact, incorporating $g_{pl}$ eventually leads to a negligible $\Pi_l(\Omega)$($\simeq$0).  

Figure \ref{fig2}b shows the fit to experimental data using the effective temperature model (see Methods). An excellent agreement is obtained for all investigated temperatures. During the fitting, the electron-phonon couplings are assumed to be independent of temperature\cite{4TM_conference_quasi_thermal, NCCO_T_independent_lambda}. In fact, the final fitting results are quite robust against the detailed shape of $\Pi(\Omega)$\cite{ETTM_Bi2212_Science} (see also Supplementary Information). As an example, the total bosonic spectra $\Pi(\Omega)$ at 35 K and the corresponding $\Pi_s(\Omega)$ and $\Pi_p(\Omega)$ ($\Pi_l(\Omega)\simeq$0) are illustrated using a histogram-like function in Fig.~\ref{fig2}c \cite{Ultrafast_review,ETTM_Bi2212_Science}. A crucial result yielded by our fitting is that the specific heat associated with $\Pi_s(\Omega)$: $C_s<0.018\,C_e$ (see Fig.~\ref{fig3}a). Such small $C_s$ evidently shows that the bosonic excitations associated with $\Pi_s(\Omega)$ have an electronic origin. This finding is also consistent well with that $T_e$ and $T_s$ get instantaneously thermalized, in contrast to the slow rising time of $T_p$ (sub-ps) and $T_l$ ($>$1 ps) in Fig.~\ref{fig2}d. Although $\Pi_s(\Omega)$ is distributed over the whole energy domain investigated, its contribution within $\sim$65 meV (the upper limit of phonon energy\cite{Gedik_2016_arXiv, NCCO_phonon_cutoff}), where different type of bosons are entangled, is significant at low temperatures. e.g. $\Pi_s/\Pi_p\simeq$0.4 for $\Omega\lesssim$65 meV at 30 K. 

In fact, our experiments show that $\Pi_s(\Omega)$ strongly depends on the temperature. Specifically, the strength of $\Pi_s(\Omega\lesssim$65 meV), represented by the red area below 65 meV in Fig.~\ref{fig2}c, increases significantly as $T$ approaches $T_c$ (see Supplementary Information). This variation can be best revealed by investigating the temperature-dependent electron-boson coupling constant: $\lambda_s=2\int\Pi_s(\Omega)/\Omega d\Omega$. As shown in Fig.~\ref{fig3}b, we clearly observe a pronounced enhancement of $\lambda_s$ below $T^*$ ($\sim$60 K), in correspondence with the strong increase of $\Pi_s(\Omega\lesssim$65 meV). Remarkably, there is a peak around $T^*$ in the specific heat $C_s$ ($<0.018\,C_e$). These results strongly indicate the emergence of electronic excitations with sufficient long correlation length below $T^*$, which have a spectral distribution within $\sim$65 meV. Indeed, previous anisotropic magnetoresistance measurements suggest that 2D AFM correlations appears below the similar temperature\cite{JinKui_AMR}. Therefore, the detected bosonic mode of electronic origin is closely related to the 2D AFM spin fluctuations, which is found to have an energy distribution mostly within $\sim$65 meV by this work. The above observations are in accordance with the remarkable change of experimental $\Delta R(t)/R$, and also explain the appearance of $A^*$ component in $\Delta R/R$, which is initially expected to arise from the onset of new scatterings by some bosonic excitations.

Assuming that each $\Pi_j(\Omega)$ ($j=s,p$) entirely contributes to the electron pairing, we are able to estimate the maximum SC transition temperature $(T_c)_j$ associated with $\lambda_j$ via the extended McMillan formula\cite{lambda_Allen_McMillan,Tc_upper_limit}: $T_c=0.83\widetilde{\Omega}_j$exp$[-1.04(1+\lambda_j)/\lambda_j]$, where ln$\widetilde{\Omega}_j=2/\lambda_j\int_{0}^{\infty}\Pi_j(\Omega)$ln$\Omega/\Omega d\Omega$. The electron-phonon coupling $\lambda_p$ is found to be $\sim$0.47 and agrees well with previous findings in the cuprates\cite{Ultrafast_review,NCCO_phonon_cutoff}. It correspondingly yields a maximum critical temperature $(T_c)_p$ of $\sim$15 K, which is far below $T_c$(=25.1 K). This implies that phonons cannot be the key paring glue alone. By contrast, the spin fluctuation mode becomes stronger with lowering temperature. As seen in Fig.~\ref{fig3}b, $\lambda_s$ at 30 K can already yield a maximum $(T_c)_s$ of 34 K, which is well above the SC transition temperature. This presents direct evidence that the AFM spin fluctuations can fully account for the electron pairing in LCCO. Given the increasing rate of $\lambda_s$ below $T^*$ nearly the same for the investigated samples, $T_c$ is expected to increase monotonically with $T^*$, and agree well with the observation in Fig.~\ref{fig1}f. Extracting $T^*$ on different optimally doped samples ($x=0.1$) gives us a new temperature-oxygen concentration ($T-\delta$) phase diagram on top of the previous $T-x$ phase diagram. The expected SC dome and the 2D AFM spin fluctuation regime are shown by the green and red regions in Fig.~\ref{fig3}c, respectively. Based on the above analysis, this phase diagram provides us evidence that stronger 2D AFM spin fluctuations will give rise to higher $T_c$.

Non-zero $\Pi_s(\Omega)$ persists to high temperatures above $T^*$, where it is dominantly distributed above $\sim$65 meV. The obtained coupling $\lambda_s$ gradually decreases with increasing temperature. For $T>T^*$, the estimated maximum $(T_c)_s$ becomes comparable with or even smaller than the $(T_c)_p$. Thus, experiments limited to high temperatures may fail to catch the decisive temperature evolution of bosonic excitations, and cannot tell which excitations are truly responsible for the SC pairing. At this stage, we cannot clarify the exact origin of the electronic excitations above $T^*$. Nevertheless, they may act as an important scattering media for the $T^{1.6}$ scaling of the resistivity by quantum criticality at the edge of superconductivity dome\cite{JinKui_PNAS,JinKui_nature}.  

Interesting phenomenon is observed when we turn to the long relaxation process in the nanosecond regime. As shown in Fig.~\ref{fig3}d, the amplitude of such long relaxation dynamics changes sign from negative to positive with decreasing temperature. Specifically, it has a negative amplitude at high temperatures, e.g. 100 K, where the long relaxation with nanosecond timescale should be mainly attributed to the thermal diffusion processes or the dissipation of low energy phonons. As the temperature gradually decreases, there appears to be a hidden positive long decay component which competes with the negative one, shifts the overall long relaxation signal towards zero at a certain temperature $T^\dagger$ (e.g. $T^\dagger\sim$16 K for sample II), and causes a positive amplitude in the nanosecond regime at lower temperatures (red dashed line in Fig.~\ref{fig3}d). We have investigated $T^\dagger$ in a set of various oxygen-tuned samples, and found that $T^\dagger$ can be either above or below $T_c$. More surprisingly, there exists an anti-correlation relation between $T^\dagger$ and $T_c$ (inset of Fig.~\ref{fig3}d). These phenomena suggest a competing order associated with $T^\dagger$ that can coexist with the superconductivity. A natural candidate of such competing order is the three-dimensional antiferromagnetism\cite{3D_AFM_competing_with_SC,2D_AFM_favor_high_Tc}. Suppression of antiferromagnetism, i.e. smaller $T^\dagger$, would lead to enhanced AFM spin fluctuations and thereby favor a higher $T_c$ for superconductivity, as evidenced by the inset of Fig.~\ref{fig3}d.

Our results provide a quantitative and unambiguous evidence in electron-doped cuprates so far that the 2D AFM spin fluctuations are the imperative glue for the SC pairing, although we cannot exclude the possibilities that other electronic excitations and phonons can also participate in the electron pairing process. Our work demonstrates that ultrafast optical spectroscopy in temperature domain not only can extract the strongest coupling associated with the electronic excitations near $T_c$ but also is capable of elucidating the origin of these excitations via their temperature-dependence. Therefore, this investigation forge a path for systematically exploring interactions between charge carriers and bosons with both electronic and phononic origins in cuprates and other correlated materials. 

\begin{methods}

\subsection{Experimental setups}
The time-resolved transient reflectivity change $\Delta R/R$ were measured using a Ti:sapphire oscillator lasing at the center wavelength of 800 nm ($\sim$1.55 eV). It has a repetition rate of 80 MHz and a pulse duration of $\sim$35 fs. For reference, experiments were also performed using a Ti:sapphire laser system with low repetition rate of 5 MHz and pulse duration of $\sim$60 fs. The pump beam, with a typical fluence of $\sim$0.3 $\mu$J/cm$^2$, directs along the normal incidence and is kept $p$-polarized. The probe beam, with a typical fluence of $\sim$0.03 $\mu$J/cm$^2$, is incident at a $\sim$10 degree angle to the sample normal and is kept $s$-polarized. Further details are given in Supplementary Information. 

\subsection{Sample syntheses}
The c-axis-oriented La$_{2-x}$Ce$_x$CuO$_{4\pm\delta}$ (LCCO, x=0.1) thin films with a thickness of 100 nm were deposited on the ($00l$)-oriented SrTiO$_3$ substrates by a pulsed laser deposition system. We used different annealing conditions to fabricate LCCO samples with various oxygen concentration ($\delta$). Since the annealing process will affect the lattice parameter ($c$) along $c$-axis and superconducting zero-resistance transition temperature ($T_c$), different $\delta$ is associated with a distinct set of $c$ and $T_c$. Detail characterization of the samples are shown in the Supplementary Information.

\subsection{Effective temperature model}
The energy transfer rate between non-equilibrium charge carriers and phonons in the two-temperature model is connected to the Eliashberg coupling spectra function $\alpha^2F(\Omega)$, and has been solved by Allen\cite{Allen_equation}. Very recently, similar theoretical frame was extended to include $I^2\chi(\Omega)$, associated with the bosonic excitations with electronic origin. The corresponding energy transfer rate is given by reference \cite{Ultrafast_review,ETTM_Bi2212_Science}:
\begin{equation}
	\label{equi1}G(\Pi_j,T_j,T_e)=\frac{6C_e}{\pi\hbar k_b^2T_e}\int_{0}^{\infty}\Pi_j(\Omega)\Omega^2[n(\Omega,T_j)-n(\Omega,T_e)]d\Omega
\end{equation}
where $j$(=$s$, $p$, $l$) represents the electronic excitations ($s$), hot phonons ($p$) and rest of the lattice ($l$), respectively. $[\alpha^2F(\Omega)]_{p,l}$ and $I^2\chi(\Omega)$ are represented by $\Pi_{p,l}(\Omega)$ and $\Pi_s(\Omega)$, respectively. A total bosonic spectral function $\Pi(\Omega)$ is defined as $\Pi(\Omega)=\Pi_p(\Omega)+\Pi_l(\Omega)+\Pi_s(\Omega)$. $T_e$ is the electronic temperature, and $T_j$ is the effective temperature characterizing each type of bosonic excitations. $n(\Omega,T)$ is the Bose-Einstein distribution and given by $n(\Omega,T)=1/(e^{\Omega/k_BT}-1)$. Here, the Fermi-Dirac distribution is assumed to build up instantaneously after photo excitation. Therefore, the energy transfer between non-equilibrium fermionic quasiparticles ($T_e$) and bosonic excitations ($T_j$), schematically shown in Fig. 2a, can be described by the effective temperature model via a set of coupled rate equations:
\begin{eqnarray}
	&\label{equi2}&\frac{\partial T_e}{\partial t}=\frac{G(\Pi_s,T_s,T_e)}{C_e}+\frac{G(\Pi_p,T_p,T_e)}{C_e}+\frac{G(\Pi_l,T_l,T_e)}{C_e}+\frac{I(t)}{C_e}\\
	&\label{equi4}&\frac{\partial T_s}{\partial t}=-\frac{G(\Pi_s,T_s,T_e)}{C_s}\\
	&\label{equi3}&\frac{\partial T_p}{\partial t}=-\frac{G(\Pi_p,T_p,T_e)}{C_p}+\frac{g_{pl}}{C_p}(T_l-T_p)\\
	&\label{equi5}&\frac{\partial T_l}{\partial t}=-\frac{G(\Pi_l,T_l,T_e)}{C_l}-\frac{g_{pl}}{C_l}(T_l-T_p)
\end{eqnarray}
where $I(t)$ is the Gaussian-like excitation source, $C_e$ or $C_j$ is the specific heat, and $g_{pl}$ describes the coupling between hot phonons and rest of the lattice due to the anharmonic decay process\cite{g_pl}. The $g_{pl}$ coupling term is neglected in previous studies\cite{ETTM_Bi2212_Science}. But our work show that it has to be included for temperature-dependent experiments. The couplings between phonons and electronic excitations are ignored \cite{ETTM_Bi2212_Science}, mainly due to the instantaneous equilibration of $T_e$ and $T_s$. In general, the specific heat of each subsystem satisfies a relation of $C_s<C_e<C_p\ll C_l$ \cite{ETTM_Bi2212_Science}. Additionally, the electron-phonon coupling function $\Pi_{p,l}(\Omega)$ is expected to be temperature-independent\cite{4TM_conference_quasi_thermal,NCCO_phonon_cutoff}. The detail fitting procedures are given in the Supplementary Information.

\end{methods}



\begin{addendum}
 \item We thank Mengkun Liu for helpful discussions. This work was funded by the Science Challenge Project of China (TZ2016004). We also acknowledge the support of the National Key Basic Research Program of China (2015CB921000), the National Natural Science Foundation of China (11474338) and the Beijing Municipal Science and Technology Project (Z161100002116011).
 \item[Author contributions] M.W., J.Q., and S.L. performed the experiments. M.W., J.X. and J.Q. analyzed the data. H.Y. and K.J. synthesized and characterized the samples. J.Q., K.J. and Y.Y. wrote the paper with contributions from all the authors. J.Q. conceived and supervised the project.  
 \item[Competing financial interests] The authors declare that they have no
 competing financial interests.
\end{addendum}

\end{document}